
\documentstyle{amsppt}
\magnification\magstep1
\NoBlackBoxes
\define\pl{\frac{1}{p}A}
\define\Sl{\!\!\!\!\slash}
\define\dsl{d\Sl}
\define\Asl{A\Sl}
\define\psl{p\Sl}


\define\g{\gamma}
\define\G{\Gamma}
\redefine\o{\omega}

\redefine\L{\Lambda}

\define\gm{\bold g}

\define\<#1,#2>{\langle #1,#2\rangle}
\define\TR{\text{tr}}
\define\dep(#1,#2){\text{det}_{#1}#2}
\define\norm(#1,#2){\parallel #1\parallel_{#2}}

\topmatter
\title  Elementary derivation of the chiral anomaly \endtitle
\date December 15, 1994\enddate
\author Edwin Langmann and Jouko Mickelsson \endauthor
\affil Theoretical Physics, Royal Institute of Technology,
S-10044, Stockholm, Sweden \endaffil
\endtopmatter

\document
\baselineskip=18pt

ABSTRACT An elementary derivation of the chiral gauge anomaly in all even
dimensions is given in terms
of noncommutative traces of pseudo-differential operators.

                     \define\Tr{\text{tr\,}}
\vskip 0.5in
The ordinary trace functional acting on  $N\times N$ matrices is always
defined and has
the fundamental property that $\Tr [A,B]=0.$ In infinite dimensions the
concept of trace becomes more subtle. Even for bounded Hilbert space
operators the trace
does not usually make sense (the sum of diagonal elements either diverges
or depends on the bases). In the case of pseudo-differential operators
there exist alternative definitions. An important trace is given by the
operator residue, [W], which picks up precisely the logarithmically
diverging part of the ordinary trace and gives it a finite value; for a
suitable class of operators this
is equal to the Dixmier trace used extensively in noncommutative geometry,
[C].

In certain  cases important in applications of operator theory to quantum
field theory there is an alternative
way to make sense of the logarithmically diverging piece. Let $d$ be the
dimension of the underlying space-time manifold.
It may happen that
even though the trace of a PSDO if formally diverging (by power counting, the
logarithmically diverging borderline case being degree = $-d$)
it has a finite value as a limit $\Lambda\to\infty$ of the ultraviolet
cut-off trace $\Tr_{\L}.$ This is the case when the leading symbol of the PSDO
is a total derivative in momentum space; in that case the cut-off trace
becomes a finite surface integral (plus possibly a finite ordinary trace of
the lower order contributions). The most notable example for this behavior
is when the PSDO can be written as a commutator.  If the degree (in momentum
space) of the commutator is precisely $-d$ then the trace depends only on the
leading symbol and is given exactly by
a surface integral. This fact has been used in calculations involving
Schwinger terms in quantum field theory, [M],[LM],[CFNW],[L]. In this note
we shall show that the symbol calculus of PSDO's combined with the use of
the noncommutative trace, as discussed above, gives a simple derivation of the
known formula for the chiral anomaly (see the reviews in [R] and [TJZW]
for earlier approaches to this problem or [BGV] for a more mathematical
discussion of the subject).

Although the use of symbol calculus of pseudodifferential operators has always
been implicit in calculations of Feynman diagrams we find that is is of great
advantage to use the symbol calculus in a systematic way. We believe that
the theory is not yet fully exploited in quantum physics and more applications
are waiting, in addition to those in [M],[LM],[CFNW],[L].

Let $D_A = \g^{\mu}(-i\partial_{\mu} + A_{\mu} P_+)$ be the \it chiral \rm
(nonselfadjoint) Dirac operator in $\Bbb R^{d},$ $d=2n$ even. Here $A$ is
a Yang-Mills vector
potential with values in a Lie algebra $\gm$ of a compact Lie group $G$ and
$P_+=\frac12 (1+\g_{d+1})$ is the chiral projection operator with
$\g_{d+1}=(-i)^n\g_1\g_2\dots\g_{2n}.$ The $2^n\times 2^n$ gamma matrices
satisfy the euclidean
anticommutation relations $\g_i\g_j+\g_j\g_i = 2\delta_{ij}.$ The Lie algebra
$\gm$ acts on the components of the Dirac field $\psi$ through
a finite-dimensional
unitary representation $\rho$ of $G.$ Thus $\psi$ takes values
in $V=\Bbb C^{2^n}
\otimes \Bbb C^N,$ where $N$ is the dimension of $\rho.$

We shall apply the symbol calculus of pseudo-differential operators and
therefore
we assume that the potential $A$ and the gauge transformations have compact
support.

A PSDO $A$ is given by its symbol $a(x,p)=\sigma(A)(x,p)$ which is a smooth
function of the
coordinates $x\in\Bbb R^{d}$ and momenta $p\in\Bbb R^{d},$ [H].
We shall consider \it the restricted class \rm of PSDO's which admit an
asymptotic expansion of the symbol as
$$a(x,p) \sim a_{k}(x,p)+a_{k-1}(x,p) +a_{k-2}(x,p)+\dots$$
where $k$ is an integer and each $a_j$ is a homogeneous matrix valued
function of the
momenta, of degree $j,$ with $|a_j|\sim |p|^j$ as $\sqrt{p_1^2 +\dots + p_d^2}=
|p|\to\infty.$
The asymptotic expansion for the product of two PSDO's
is given by the formula
$$a*b \sim \sum \frac{(-i)^{|m|}}{m!} \left[(\partial_p)^m a(x,p)\right]
\left[ (\partial_x)^m
b(x,p)\right],\tag1$$
where the sum is over all sets of nonnegative integers $m=(m_1,\dots,m_{d}),$
$|m|=m_1+\dots +m_{d},$ ${\partial_x}^m=(\frac{\partial}{\partial x_1})
^{m_1}\dots(\frac{\partial}{\partial x_d})^{m_{d}},$ etc., and
$m!=m_1!\dots m_{d}!$.

The trace (if it exists) of a matrix valued PSDO $A$ is given as
$$\Tr \, A= \frac{1}{(2\pi)^{d}} \int \Tr \, a(x,p) dx\, dp.\tag2$$
The trace exists if the symbol $a(x,p)$ behaves asymptotically $|p|^{-d-
\epsilon}$ for a positive $\epsilon,$ assuming that $a(x,p)$ has compact
support in $x$ (this condition is satisfied in the cases of interest to us).
The borderline case $\epsilon=0$ is
important to us: in that case the trace diverges logarithmically \it
except when the leading symbol $a_{-d}(x,p)$ is a total derivative in
momentum space. \rm If it is a total derivative the trace can be written as
a (finite) surface integral of some symbol which is of degree $-2n+1.$
For the restricted class of PSDO's $A$ with a symbol $a$ of degree $k$ we can
expand the
\it cut-off trace \rm $\text{tr}_{\L} \,A$ in powers of the momentum
space cut-off $\L,$
$$\align \Tr_{\L}\,A&= \frac{1}{(2\pi)^{d}} \int_{|p|\leq \L} dp
\int dx\,\Tr a(x,p)\\
&= c_{k+2n} \L^{k+2n} + c_{k+d-1} \L^{k+d-1} + \dots c_{log} \log \L
+c_0 \L^0 +c_{-1} \L^{-1} +\dots\tag3 \endalign$$
The logarithmic term arises from the trace of the degree $-d$ piece in the
symbol and $c_{log}(A)$ is proportional to the operator residue of $A.$
The regularized trace of a PSDO is then defined as
$$\text{TR}\,A= c_0(A)$$
in the above expansion. If deg$A < -d$ then the regularized trace is equal to
the ordinary trace $\Tr A.$

The degree of $a*b$ is  equal to the sum of degrees of $a$ and $b.$
The leading term in $a*b$ is just the matrix product $ab$ of the symbols.
If deg$a$ + deg$b =p$ the highest order term which contributes to the
trace $\Tr[A,B]$ is of degree $p-1$ since the finite-dimensional commutator
traces in $V$ vanish identically.

Our calculation of the anomaly is based on the fact that we can write
the gauge variation of the effective action as a sum of traces of commutators.
For bounded operators $A,B$ the trace $\Tr[A,B]$ is stable with respect to
perturbations of $A,B$ by trace class operators. Therefore, it depends only
on the asymptotic expansion of the symbols of $A$ and $B.$ This is because two
PSDO's with the same asymptotic expansion can differ only by a so-called
infinitely smoothing operator. An infinitely smoothing operator is always
trace-class (for compactly supported symbols in $x$ space).

\redefine\TR{\text{TR}\,}

The effective action for chiral fermions coupled to external Yang-Mills field
is defined as (the logarithm of) $\text{det}_{reg}
(D_A)$ where 'reg' means that some sort of regularization is needed for
the convergence of the infinite-dimensional determinant. We shall use the
following regularization. Formally, the logarithm of a determinant is equal to
the trace of the logarithm of the operator. Thus we define tentatively
$$\log\, \text{det}_{reg}(D_A) = \TR\,\log(D_A).$$
Of course, the right-hand-side is ill-defined when $D_A$ is not invertible
and for invertible operators one has to fix the phase of the logarithm
somehow. It will turn out that for the computation of chiral anomaly the
form of the 'infra-red' regularization of the Dirac operator is inessential
(since the anomaly will be a surface integral in momentum space). In the
following discussion, leading to the definition (5) of the effective action,
we shall therefore not be specific about the choice of the infrared
regularization; we shall return to this problem later.

The operator log$(D_A)$ is not a classical pseudo-differential operator,
but we can write log$(D_A) = \text{log}(D_0) +W_A,$ where $W_A$ is a classical
pseudodifferential operator of degree zero. We can then drop an (infinite)
constant from the definition of the effective action by declaring it to be
equal to
\redefine\log{\text{log}}  $\TR\, W_A.$ The operator $W_A$ is
very complicated function of $A$ because
$$\log(ST) =\log(S) +\log(T) +F(S,T)$$
where $F(S,T)=\frac12[\log(S),\log(T)] +$ higher commutators. When computing
the regularized trace one is not allowed to drop the commutator terms because
in general $\TR [S,T] \neq 0.$ However, only a finite number of commutators
are relevant, giving a contribution to the effective action which is a finite
differential polynomial. This is seen as follows. In our case $S=D_0$ and
$T={D_0}^{-1}D_A=1 +{D_0}^{-1} {\Asl}_+$ where ${\Asl}_+ =\g^{\mu}
A_{\mu} P_+.$ Thus
\define\DA{{D_0}^{-1} {\Asl}_+}
$$\log(T) =  \DA -\frac12 (\DA)^2 +\frac13 (\DA)^3+ \dots.\tag4$$
It follows that there are only a finite number of terms in the expansion of the
logarithm which are of degree $> -d$ and which can contribute to the
trace of the
commutator. Furthermore, for each term we need to compute derivatives in
momentum space (for the asymptotic expansion (1)) up to a finite degree,
because
each derivative decreases the degree by one unit. Finally, in the multiple
commutator expansion for $F(S,T)$ we need to take only a finite number of
commutators with $T$ because the degree of $\log(T)$ is $-1.$

The conclusion is that
$$\TR W_A =\TR \log(1+\DA) + P(A)$$
where $P(A)$ is a finite differential polynomial in $A.$  After performing this
polynomial renormalization we define
$$S_{eff}(A)= \TR \log(1+\DA).\tag5$$
This definition gives also a resolution to the phase problem for the logarithm.
For small potentials $A,$ after fixing an infrared regularization for $D_0$
(e.g. $D_0^{-1} \to (D_0 +i\epsilon)^{-1}$) one can expand $\log(1+\DA)$ as a
convergent power series in $\DA.$ A similar expansion can be performed around
an arbitrary background potential $B$ by replacing $D_0\to D_B.$

The chiral anomaly is defined as $\o(X;A)=\Cal L_X S_{eff}(A),$ where
$\Cal L_X$
denotes the Lie derivative in the direction of an infinitesimal gauge
transformation
$X:\Bbb R^{2n} \to \gm,$ $(\Cal L_X f)(A)= \frac{d} {dt} f(-i e^{-itX}
d e^{itX} +e^{-itX}Ae^{itX})|_{t=0}.$
We shall show below that the anomaly can be written as
a regularized trace of commutators of PSDO's which then leads automatically
to a finite differential polynomial expression for the anomaly.

Using $-i \Cal L_X \Asl = -i\dsl X +[\Asl,X],$ where
$\dsl X=\gamma^{\mu}\partial_{\mu}
X,$ $\Asl=\g^{\mu} A_{\mu},$ etc.,
we get a symbol formula for $\o(X;A),$
$$\align \o(X;A) = \sum_{k=1}^{\infty} \frac{(-1)^{k-1}}{k}
\sum_{j=0}^{k-1}\TR (\frac{1}{ D_0}{\Asl}_+)^j \left\{\frac{1}{ D_0}
\dsl X P_+ \right.\\
&\left.+\frac{1}{ D_0}i[\Asl,X] P_+\right\} (\frac{1}{ D_0} {\Asl}_+)^{k-j-1}.
\tag6\endalign$$
(In computing the traces a fixed infrared regularization $1/ D_0 \to 1/( D_0+
i\epsilon)$ is used.)
On the other hand,
$$\align &\TR[\log(1+\frac{1}{ D_0} {\Asl}_+),iX]=
\sum_{k=1}^{\infty} \frac{(-1)^{k-1}}{k} \sum_{j=0}^{k-1}\\
& \TR(\frac{1}{ D_0} {\Asl}_+)^j
\left(-\frac{1}{ D_0}  \dsl X\frac{1}{ D_0} {\Asl}_+ + \frac{i}{ D_0}
[\Asl,X]P_+\right)(\frac{1}{ D_0}{\Asl}_+)^{k-j-1}
.\tag7\endalign$$
Disregarding commutator terms this expression is the same as the above formula
for $\o(X;A).$ More precisely,
$$\align \o(X;A) &= \TR [\log(1+\frac{1}{ D_0} \Asl),iXP_+]\\
&+\sum_{k=1}^{\infty} \frac{(-1)^{k-1}}{k(k+1)} \sum_{j=0}^{k-1} \TR
\left[(\frac{1}{ D_0} \Asl)^j(\frac{1}{ D_0} \dsl X) , (\frac{1}{ D_0}
\Asl)^{k-j}P_+\right]. \tag8\endalign$$
We have uses the projection operator property ${P_+}^2=P_+$ and the fact that
$P_+$ commutes with a product of a pair of $\g$ matrices. Since the anomaly
is a sum of traces of commutators it really depends only on the asymptotic
expansion of the operators involved.

Actually, there is a delicate point in the above argument which shows the
difference between the terms arising from the $\g_{d+1}$ part (the nontrivial
part of the anomaly) or from $'1'$ in the projection $P_+=\frac12
(1+\g_{d+1}).$ Remember that we have to make an infrared regularization
\define\pe{\frac{1}{ D_0+i\epsilon}}  $\frac{1}{ D_0}\to\pe$
in order to make sense of the traces. But the regularized operator $\pe \Asl$
does
not commute with $\g_{d+1}$ which prevents us of writing the gauge variation
of the operator $\log(1+\pe \Asl P_+)$ exactly as a sum of commutators; but
this
does not have an effect on the computation involving traces. The reason is
that the nonzero commutators $[P_+,\pe \Asl]$ affect only terms arising from
the
commutator $[\log(1+\pe \Asl P_+),iX].$ The ultraviolet regularized trace of
the
commutator depends only on the degree $-d$ part of the asymptotic expansion
of the commutator. Thus the choice of the infra-red regularization does not
have
any effect on the result and can be ignored (see the explicit computations
below); the remaining traces will be surface integrals in the momentum space.

A basic formula for the $\g$-matrix algebra which we need several times is
$$\Tr \g_{i_1}\g_{i_2}\dots \g_{i_p}\g_{2n+1} =\cases 0, \text{ if } p<2n \\
           (2i)^n \epsilon_{i_1 i_2\dots i_{2n}}, \text{ if } p=2n.
           \endcases \tag9$$
We shall also use $\sigma(D_0)(x,p)=\psl=\g^{\mu}p_{\mu}.$

\bf The case $d=2.$ \rm Since now $\TR[A,B]=0$ when $deg(A)+deg(B) \geq 2,$ the
only nonvanishing term is
$$\o(X;A)= \TR [\frac{1}{ D_0} \Asl,iXP_+].\tag10$$
The form of the infrared regularization at $p=0$ is irrelevant, since the
trace of
the commutator will be a integral of a total derivative in momentum space.
$$\align \o(X;A)&= \frac{1}{(2\pi)^2}\int dx \int dp (-i) \Tr
\frac{\partial}{\partial
p_{\mu}}(\frac{1}{ \psl})\Asl \frac{\partial}{\partial x_{\mu}}i X P_+\\
&= \frac{1}{(2\pi)^2} \int dx \int_{|p|=\infty}dp\, \Tr \frac{p_{\mu}}{|p|}
\frac{ \psl}{|p|^2}\Asl \partial_{\mu} X P_+\\
&= \frac{1}{4\pi}\int dx\, \Tr \g^{\mu}\g^{\nu} A_{\nu} \partial_{\mu} X
\frac12(1+\g_3)\\
&= \frac{i}{4\pi} \int dx \epsilon^{\mu\nu} \Tr A_{\mu}\partial_{\nu} X
+ \frac{1}{4\pi} \int dx\, \Tr A_{\mu} \partial_{\mu}X.\tag11\endalign $$

The real part (the second term) is a coboundary of the local functional
$f(A)= \frac{1}{8\pi} \int dx \Tr |A|^2$ and thus a trivial 1-cocycle.
Actually, the real part must be a trivial cocycle in any space-time dimension
on
cohomological grounds. Note also that the real part is coming from the term 1
in $P_+=\frac12 (1+\g_{2n+1}).$ This term is the gauge anomaly for the
nonchiral
determinant $\frac12\TR\,\log(1+\frac{1}{D_0} \Asl).$ We shall show later
explicitly
that really $\Cal L_X \TR \log(1+\pe \Asl)=\Cal L_X P_d(A)$ where $P_d$
is a local differential polynomial in $A$ in any dimension $d.$

\bf The case $d=4.$ \rm Now the relevant terms contributing to the commutator
trace are
$$\align \o(X;A)&= -\frac12 \text{TR} [(\frac{1}{ D_0}\Asl)^2,iXP_+]
           +\frac13 \text{TR} [(\frac{1}{ D_0}\Asl)^3, iXP_+]  \\
          & -\frac16 \text{TR} [\frac{1}{ D_0} \Asl\frac{1}{ D_0} \dsl X,
          \frac{1}{D_0}\Asl P_+]
           -\frac16 \text{TR} [\frac{1}{ D_0} \dsl X,
           (\frac{1}{ D_0}\Asl)^2P_+].\endalign$$
The actual computation of the various commutators and traces is a
straight-forward
application of (1) for the product of PSDO's and the $\g$ matrix algebra (9).
The result for the imaginary part is
$$\o_{\g}(X;A)= \frac{-1}{24 \pi^2 } \int \Tr dX (-idA A +\frac12 A^3)
\tag12$$
where the exterior algebra notation has been used for the integrand. The real
part is again trivial and can be written as a coboundary of a suitable
differential polynomial in $A.$

\bf The general case \rm
                               \redefine\pl{\frac{1}{D_0}\Asl}

First we regroup the terms in (8). We can write
$$\align -i\o(X;A) &= \sum_{m=1}^{\infty} \frac{(-1)^{m-1}}{m(m+1)}
\sum_{j=0}^{m}
\TR \left\{[(\pl)^{j}, X (\pl)^{m-j} P_+] -\right.\\
&\left. -[(\pl)^j X, (\pl)^{m-j} P_+]
+[(\pl)^j \frac{1}{ D_0}[ D_0,X], (\pl)^{m-j}P_+]\right\}.\tag13\endalign $$
The right-hand side can be written in a more compact form, involving the
inverse
of the Dirac operator $D_A,$  by the following
trick.   Using the identity
$$\frac{-1}{m(m+1)} = \int_0^1 t^m (1-\frac{1}{t}) dt$$
we get
$$\o(X;A) = \int_0^1 \nu(X;tA) \frac{t-1}{t} dt \tag14$$
with
$$\align -i\nu(X;A) &= \sum_{m=0}^{\infty}\sum_{j=0}^{m}(-1)^m \TR \left\{
[(\pl)^j,X(\pl)^{m-j} P_+]-\right.\\
&\left.- [(\pl)^j X,(\pl)^{m-j}P_+] +[(\pl)^j \frac{1}{ D_0}[
D_0,X],(\pl)^{m-j}
P_+]\right\}    \tag15\endalign$$
Using a geometric series expansion for $(D_A)^{-1} =( D_0+\Asl)^{-1}$
we can write
$$-i\nu(X;A) = \TR \{ [(D_A)^{-1} D_0, X (D_A)^{-1} D_0 P_+]
-[(D_A)^{-1} X D_0,(D_A)^{-1} D_0 P_+]\}.\tag16$$
The symbol of the PSDO on the right-hand side can be expanded in a geometric
series using
$$\sigma(D_A^{-1} B)(x,p)= ( D_0-i\g^{\mu}\frac{\partial}{\partial x_{\mu}}
+\Asl(x))^{-1} \sigma(B) (x,p)$$
where $B$ is a PSDO.
The nontrivial part of the anomaly arises from the $\g_{d+1}$ term in $P_+$
(see the discussion below on the real part of the anomaly). We shall denote
by $\nu_{\g}$ the sum of $\g_{d+1}$ terms in $\nu.$
                        \define\pp{\frac{1}{ D_0}}
The trace of the commutators is again a boundary integral in momentum space
of terms with symbols decaying like $|p|^{-d}$ as $|p|\to \infty.$
Expanding (16) as geometric series in momentum space,
\redefine\pp{\frac{1}{\psl}}
$$\align \nu_{\g}(X;A) & = \frac{1}{2(2\pi)^d}\int dp \delta(\L-|p|)
\int dx\,\frac{p_k}{|p|}
\sum_{m+r=d-1} \Tr (\pp \g^{i_1} \dots \pp \g^{i_m}\pp\g^{j_1}\dots
\pp\g^{j_r}\g_{d+1})\times \\
& \times \Tr \left\{ (a_{i_1\dots i_m}\cdot 1) \partial_{k} (X a_{j_1\dots j_r}
\cdot 1)
-(a_{i_1\dots i_m}\cdot X)\partial_k (a_{j_1\dots j_r}\cdot 1) \right\}
\tag17\endalign$$
where
$$a_{i_1\dots i_m} = (-i\partial_{i_1} + A_{i_1})\dots (-i \partial_{i_m}
+A_{i_m}).$$
Taking into account that the nonzero contribution to the trace of the
commutators comes from degree $=-d$ terms in momentum space and using the
property (9) of the traces of gamma matrices, we obtain
$$\align \nu_{\g}(X;A) &= \frac{1}{2d(2\pi)^d}\text{vol}(S^{d-1}) (2i)^{d/2}
(-1)^{(d-2)/2} \sum_{m+r=d-1} (-1)^m
\epsilon_{i_1\dots i_m k j_1\dots j_r}\\
& \int dx\, \Tr\left\{ (a_{i_1\dots
i_m}\cdot 1)\partial_k( X a_{j_1\dots j_r}\cdot 1)-(a_{i_1\dots i_m}\cdot
X)\partial_k(
a_{j_1\dots j_r}\cdot 1) \right\}\,\tag18\endalign$$
where vol$(S^{d-1})= 2\pi^n/\G(n).$
The antisymmetrized coefficients $a_m(A)=\epsilon_{i_1\dots i_m} a_{i_1\dots
i_m}$ are evaluated using $F=-idA+A^2$ and the
Bianchi identity $-idF +[A,F]=0.$ We get
$$a_m(A)\cdot Z = \cases F^{m/2}Z  \text{ when $m$ even }\\
                  F^{(m-1)/2} (-idZ+AZ) \text{ when $m$ odd }
                  \endcases$$
where $Z$ is an arbitrary operator and the products are exterior products of
differential forms.

Inserting this to (18) and dropping integrals of total derivatives
one gets
$$\nu_{\g}(X;A) \propto \sum_k \int dx\,\Tr dX F^k A F^{n-k-1}.$$
Using the symmetrized trace
$$\text{Str}(T_1,\dots ,T_m)
= \frac{1}{m!} \sum_{perm} \Tr T_{i_1} \dots T_{i_m}$$
the final formula for $\nu_{\g}$ can be compactly written as
$$\nu_{\gamma}(X;A) = -\frac{(-i)^{n}}{(n-1)! (2\pi)^n}
\int dx\, \text{Str}\, dX F^{n-1} A.\tag19$$
In the above formula one has to insert $dX,A$ and $(d-2)/2$ copies of $F$ as
arguments.

Summarizing, we get for the $\g_{d+1}$ part of the anomaly,
$$\o_{\g}(X;A) = - \frac{(-i)^{n}}{(n-1)! (2\pi)^n} \int_0^1 dt \frac{t-1}{t}
\int dx\, \text{Str}  (dX A_t F_t^{(d-2)/2})\tag20$$
where $A_t=tA$ and $F_t= -it dA +t^2 A^2.$

\vskip 0.3in
\bf The real part of the anomaly \rm

The term in $\nu(A;X)$ due to the unit matrix in $P_+=\frac12(1+\g_{d+1})$
is equal to the gauge variation of $\frac12 \TR \text{log}(1+\frac{1}{D_0+
i\epsilon} \Asl).$
We have an exact operator equation

$$\align \Cal L_X \text{log}(1+\frac{1}{D_0+i\epsilon} \Asl) &=
[\log(1+\frac{1}{D_0+i\epsilon} \Asl),iX]\\
&+\sum_{k=1}^{\infty} \frac{(-1)^{k-1}}{k(k+1)} \sum_{j=0}^{k-1}
\left[(\frac{1}{D_0+i\epsilon} \Asl)^j(\frac{1}{D_0+i\epsilon}  \dsl X) ,
(\frac{1}{D_0+i\epsilon} \Asl)^{k-j}\right].
\tag21\endalign$$

In contrast to the  chiral case the derivation of this equation does
not involve the
problem that $\gamma_{d+1}$ does not anticommute with
$\frac{1}{D_0+i\epsilon}.$

The $\Lambda$ finite part of the cut-off trace of the right-hand side depends
only on the projection $\pi_{-d}$ (because it is a sum of commutators)
of the asymptotic expression to the degree $-d$
part. On the other hand, a gauge variation commutes with the degree operator
and therefore
$$\align \text{Re}\, \omega(A;X)& = \Cal L_X \text{tr}_{\Lambda} \pi_{-d}
     \frac12 \text{log}(1+\frac{1}{D_0+i\epsilon} \Asl)\\
                       & = \Cal L_X P_d(A)\endalign$$
Clearly $P_d(A)$ is a local differential polynomial in $A,$ of degree
$d.$ In the case $d=2$ it is equal
to $\frac{1}{8\pi}\int \Tr |A|^2 dx.$
This argument is not valid for the imaginary part. We  do \it  not \rm have an
exact operator equation like
$$\Cal L_X \text{log}(1+\frac{1}{D_0+i\epsilon} \Asl P_+) = \sum [R_i,T_i]$$
for some operators $R_i,T_i.$

Actually, one never needs an explicite expression for $P_d(A)$ since in the
modified effective action (which has gauge invariant real part) precisely
these terms are subtracted. For those operators for which the logarithms are
defined we can write
$$\hat S_{eff}(A) = \TR \left[\text{log}(1+\frac{1}{D_0+i\epsilon} \Asl P_+)
-\frac12 \pi_{-d}
\text{log} (1+\frac{1}{D_0+i\epsilon} \Asl)\right].\tag22$$
This is our final formula for the effective action with gauge invariant
real part. There is an alternative way to define the trace without using
the cut-off mechanism. Since the operator $\pl$ is of degree $-1$ the
powers $(\pl)^k$ are trace-class when $k>d.$ Therefore the operator
$\text{log}(1+\pl P_+) -\sum_{k=1}^{d} \frac{(-1)^{k-1}}{k} (\pl)^k P_+$
has a finite trace without any cut-off. However, we have not performed
the anomaly calculation for this form of the effective action although we
expect to get an equivalent result.

\vskip 0.3in
ACKNOWLEDGEMENTS We want to thank G. Ferretti and S. Rajeev for many
discussions on anomalies.

\newpage
REFERENCES

\vskip 0.3in
[BGV] N. Berline, E. Getzler, and M. Vergne: \it Heat kernels and
Dirac operators. \rm Springer-Verlag, Berlin (1992)

[C] A. Connes: \it Noncommutative geometry. \rm Academic Press (1994)

[CFNW] M. Cederwall, G. Ferretti, B. Nilsson, A. Westerberg: Schwinger
terms and cohomology of pseudodifferential operators. {\tt hep-th/9410016}

[H] L. H\"ormander: \it The analysis of linear partial differential
operators III. \rm Springer-Verlag, Berlin (1985)

[L] E. Langmann: Non-commutative integration calculus. {\tt hep-th/9501092}

[LM] E. Langmann and J. Mickelsson: $(3+1)$-dimensional Schwinger terms
and non-commutative geometry. Phys. Lett. \bf B 338, \rm p. 241 (1994)

[M] J. Mickelsson: Wodzicki residue and anomalies of current algebras.
   {\tt hep-th/9404093}. In: \it Integrable Models and Strings, \rm ed. by
   A. Alekseev, A. Hietam\"aki, K. Huitu, A. Morozov, and A. Niemi.
   Lecture Notes in Physics \bf 436, \rm Springer-Verlag, Berlin (1994).

[R] R. Rennie: Geometry and topology of chiral anomalies in gauge
theories. Adv. Phys. \bf 39, \rm p.617 (1990)

[TJZW] S.B. Treiman, R. Jackiw, B. Zumino, and E. Witten: \it Current algebra
and anomalies. \rm Princeton University Press, Princeton (1985)

[W] M. Wodzicki: Noncommutative residue. In: Lecture Notes in Mathematics
\bf 1289, \rm ed. by Yu. Manin, Springer-Verlag, Berlin (1985)

\enddocument